\documentclass[11pt,a4paper]{article}
\usepackage{graphicx}
\usepackage{enumerate}
\usepackage{amsmath}
\usepackage{lineno}
\usepackage[labelfont=bf, labelsep=period]{caption}

\newcommand{\figref}[1]{\figurename~\ref{#1}}
\usepackage{natbib}
\author{Michel Crucifix \\ Georges Lema\^\i tre Center for Earth and Climate Research \\ 
Earth and Life Institute \\ Universit\'e catholique de Louvain \\ Louvain-la-Neuve \\ Belgium \\ \textsf{michel.crucifix@uclouvain.be}}
\title{How can a glacial inception be predicted? \\ \normalsize{ 
Contribution to the special issue of the Holocene on the Early Anthropogenic Hypothesis}}
\date{Published in 'The Holocene',  vol. 21, pp. 831-842, 2011, doi:10.1177/0959683610394883 \\ 
note : minor modifications may have been introduced at the proof-editing stage}
\begin{document}
\maketitle
\begin{abstract}
The Early Anthropogenic Hypothesis considers that greenhouse gas concentrations should have declined during the Holocene in absence of humankind activity, leading to glacial inception around the present. 
It partly relies on the fact that present levels of northern summer incoming solar radiation are close to those that, in the past, preceded a glacial inception phenomenon, associated to declines in greenhouse gas concentrations. However, experiments with various  numerical models of glacial cycles show that next glacial inception may  still
be delayed by several ten thousands of years, even with the assumption of greenhouse gas concentration declines during the Holocene.  Furthermore, as we show here, conceptual models designed to capture the gross dynamics of the climate system as a whole suggest also that small disturbances may sometimes cause substantial delays in glacial events, causing a fair level of unpredictability on ice age dynamics. This suggests the need of a validated mathematical description of the climate system dynamics that allows us to quantify uncertainties on predictions. 
Here, it is  proposed to organise our knowledge about the physics and dynamics of glacial cycles through a Bayesian inference network. Constraints on the physics and dynamics of climate can be encapsulated into a stochastic dynamical system. These constraints include, in particular, estimates of the sensitivity of the components of climate to external forcings, inferred from plans of experiments with large simulators of the atmosphere, oceans and ice sheets. 
On the other hand,  palaeoclimate observations are accounted for through a process of parameter calibration. We discuss promises and challenges raised by this programme.
\end{abstract}
\section{Introduction}
The Early Anthropogenic Hypothesis considers  that the natural  evolution of climate consisted in a decline in greenhouse gas concentrations throughout the Holocene, leading today to conditions favourable to accumulation of ice in the Northern Hemisphere (\citet{Ruddiman11aa} and references therein).  This hypothesis supposes an important premise: it is possible to predict the slow evolution of climate several millennia ahead. Indeed, suppose that climatologist lived 8,000 years ago. What the Early Anthropogenic Hypothesis says is that a forecast for the 8,000 years to come made by this early climatologists would have been a decline in greenhouse gas concentrations and ultimately, glacial inception. 

Throughout this paper we will consider that this premise should not be taken for granted. The general problem of predicting the evolution of the climate system several millennia ahead may be tackled from different perspectives. 
One method of prediction consists in seeking episodes in the past during which the climatic conditions and forcings were analogous to those of the Early Holocene, and observe the subsequent climate evolution. The other method consists in using models, generally numerical models, which account for known physical constraints on the dynamics of the climate system. 
The two methods are briefly reviewed and discussed in the next section of this article. 

Most investigators generally appreciate that  the two methods need, in practice, to be combined. 
 The question, addressed in section 3, is how observations and models may be combined in a way that is  transparent, optimal, and avoids the dangers of a circular reasoning. The methodology proposed here consists in (a) formulating stochastic dynamical systems capturing palaeoclimate dynamics. These may be viewed as generators of palaeoclimatic histories, designed to encapsulate information inferred from experiments with large numerical models and/or other hypotheses about climate system dynamics;  (b) calibrate the parameters and initial conditions of these dynamical systems on palaeoclimate observations. We explain why the Bayesian statistical framework is adapted to this programme. 



\section{Traditional approaches to predict the slow evolution of climate}
\subsection{The analogues method and its limitations}

Among the solutions envisioned by Edwar Lorenz \citeyearpar{lorenz69aa} to forecast weather, was an empirical approach also known the  `analogues method'. Considering that  similar causes should lead to similar effects\textit{ at least within a finite time horizon}, weather predictions may be attempted by seeking \textit{analogous} synoptic situations to today's in the archives of meteorological services. Likewise, predicting the natural climate evolution during the Holocene may be addressed by identifying in palaeoclimate archives situations for which the state of climate and the forcing conditions were similar to those experienced during the Holocene. 

At these times scales climate is influenced by the slow changes in  incoming solar radiation, and the latter is a function of the Earth orbital elements (characterised by the eccentricity $e$ and the true solar longitude of the perihelion $\varpi$) and the tilt---or obliquity---of the equator on the ecliptic, denoted $\varepsilon$. These three elements have all specific and well-known effects on the seasonal and spatial distributions of incoming solar radiation (insolation). They also have their own periodicities: You never encounter twice a same orbital configuration (see, on this topic, the early works of \citet{Berger77ber2} and \cite{berger79cim}). 

Palaeoclimatologists therefore tend to consider specific measures of insolation, supposed to summarise in one quantity the effect of insolation on climate, and look at past times where that specific measure of insolation was the same as during the Holocene.
However, the choice of an `analog'  depends on how these measures are constructed \citep{Loutre08aa}.

\begin{figure}[ht]
\begin{center}
\includegraphics{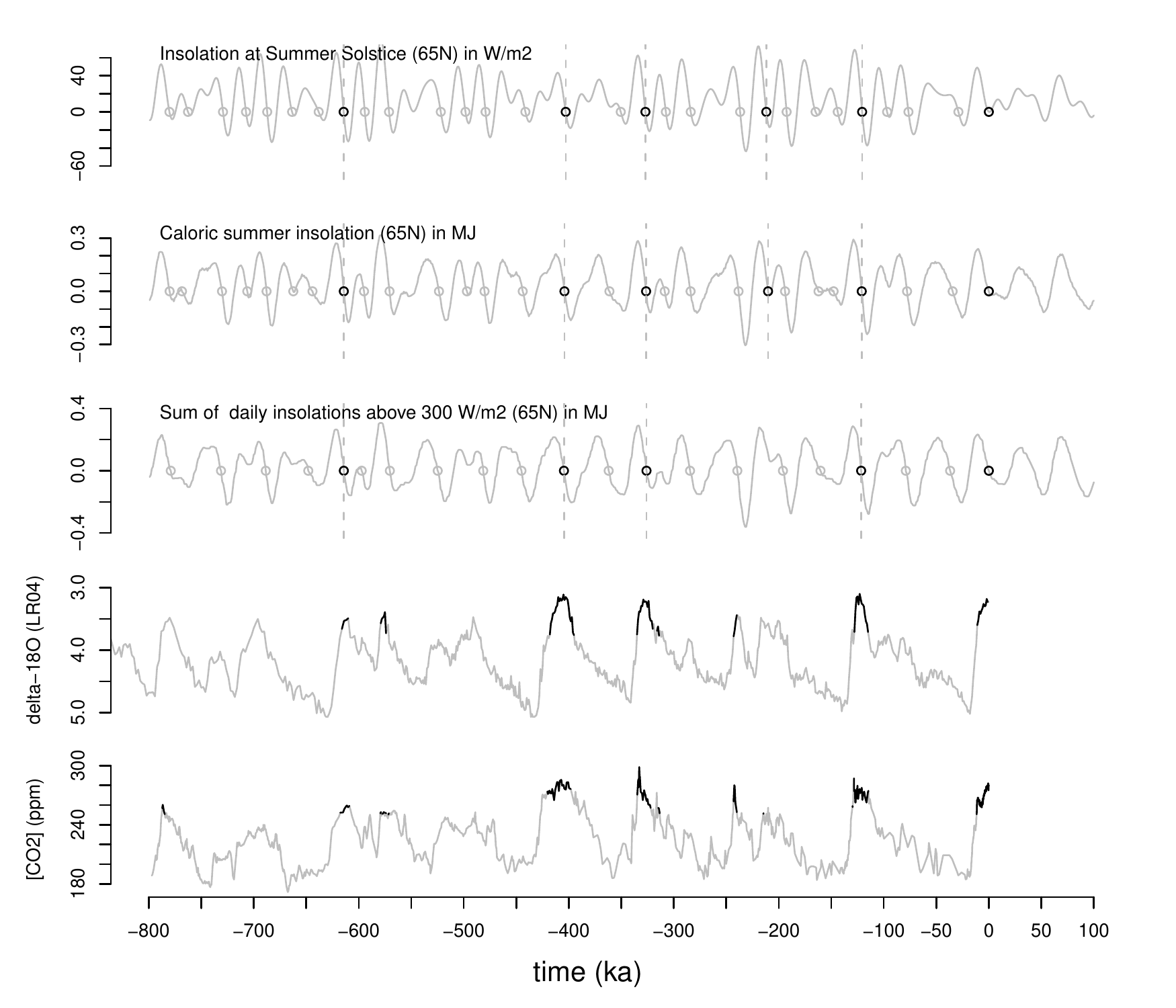}
\end{center}
\caption{
  Three measures of insolation covering the period $-800$\,ka to $+100$\,ka (departure to the present) calculated using the \cite{berger78} solution, aligned with a stack of 57 benthic foraminifera $\delta^{18}O$ \citep{lisiecki05lr04} and a reconstruction of atmospheric CO$_2$ concentration \citep{Luethi08aa,petit99}. Circles mark insolations equal to the present. Times corresponding to CO$_2$ concentrations above 250~ppm and benthic $\delta^{18}O$ below 3.8 per mil are highlighted to subjectively mark interglacial conditions. 
\label{fig:insol}}
\end{figure}

To illustrate this point consider three measures of insolation classically referred to in the tradition of Milankovitch's theory (\figref{fig:insol}): daily mean insolation at the summer solstice (21th June), calculated according to \cite{berger78}; (b) insolation integrated over the caloric summer (the 180 days of highest insolation, assuming a 360-day year) \citep{Berger1978139}---this measure is favoured by \citet{ruddiman07rge}--- (c) insolation integrated over all days for which insolation exceeds 300 W/m$^2$ (after a proposal by \cite{Huybers08aa}). All are calculated here for a latitude of 65$^\circ$ N. 
It is easily shown that these measures of insolation are  very well approached by a linear combination of climatic precession ($e\sin(\varpi)$) and obliquity ($\varepsilon$);  they are here 
ordered by increasing influence of obliquity vs climatic precession. 

Circles on \figref{fig:insol} indicate times at which the different insolation measures equal their present-day values. Call these `insolation matches'. Those occurring during a climate environment analogous to the present Holocene (arbitrarily defined as a CO$_2$ concentration larger than 250~ppm and benthic $\delta^{18}O$ less than 3.8~per mil) are  highlighted by vertical bars. 

The timing of insolation matches differ by about 1~ka (thousands of years) depending on which insolation is considered: for example the time of insolation match at marine isotopic stage (MIS) 11 ranges between $-402.9$~ka (daily mean at solstice) and $-404.5$~ka (Huybers' thresholded sum).  
In general, though, insolation levels similar to the present day slightly precede or coincide with times during which ice volume markedly increased. 
Bearing in mind the caveats introduced by uncertainties on the chronology of the palaeoclimate records, this  observation suggests that today's insolation levels are  close to those that prevailed at the end of previous interglacial periods. However, observe also that  present-day  insolations are  at a minimum on their secular course (daily mean solstice) or  close to it (season-integrated insolations). This remark invites us to be prudent about the prediction of a natural end  to the Holocene around the present, again assuming no anthropogenic perturbation. 


More crucial differences among insolation measures lie in their dynamics.  
Climate does not respond instantaneously to the astronomical forcing. Ice build-up and ocean alkalinity adjustments involve time scales of the order of 10,000 years at least. 
Predicting the evolution of climate therefore  requires one to consider the  history of astronomical forcing  over long enough a time-window,
corresponding to the time needed by the different climate system components to adjust dynamically to that specific astronomical history.  \cite{ruddiman07rge} and \cite{Loutre08aa} noted that the astronomical history preceding $-325$~ka (marine isotopic stage 9) admittedly looks as a reasonable candidate to the last 20~ka by reference to season-integrated insolations (consider the 'shapes' of the caloric summer insolation before  $-325$~ka); though, it is not straightforward to demonstrate that this really is the `best possible analog' quantitatively, in a way that is convincingly robust and objective (Loutre, pers. comm. and further tests made for this article). 

There are further caveats to a straightforward use of insolation analogues  to predict climate change. 
First, which insolation (daily mean solstice or seasonal integration) is the best predictor of climate change may depend on the level of glaciation \citep{Rohling201097}. 
Second, the analogues method supposes as a premise that one single possible climate history may correspond to a given history of astronomical forcing. In the formal language of dynamical system theory this property is called \textit{general synchronisation} of climate on astronomical forcing (see \cite{PhysRevE.51.980} for a formal definition). It is shown in the next section that this might  not be the case. 
Third, climate cycles evidently changed in shape and variance over the last million years due to slow changes in the environment. 
For example, the mid-Pleistocene revolution, when glacial cycles shifted from a period of 40~ka to a period of the order of 100~ka is explained by \citet{saltzman90sm} as a bifurcation related to a slow, tectonically-driven decline in background levels of CO$_2$. \citet{Clark06aa} review a number of alternative explanations. The mid-Br\"unhes transition about 400~ka ago, after which interglacial periods became characterised by higher greenhouse gas concentrations than before this transition, is still unexplained (see the review of \cite{Tzedakis09aa}).
Consequently,  one cannot be assured that the levels of insolation  leading to glacial inception determined on the basis of an event 400~ka ago still hold valid today. 

\subsection{Using climate simulators to predict the glacial inception}
The other approach to the problem of predicting an ice age proposed by Edwar Lorenz relies on a mathematical model of the Earth's climate. In a visionary paper entitled ``climate modelling as a mathematical problem'' \citep{Lorenz70aa}, he expresses the hope that \textit{``in the 21st century''}, large mathematical models accounting for \textit{``every feature of the atmosphere and its environment''} as well as ocean circulation and vegetation could be run and predict ice ages. 

The `mathematical' models envisioned by Lorenz are known today as `general circulation models' or `global climate models'. 
They reproduce many features of the climate system visible at the global scale (the ocean circulation, modes of variability such as El-Ni\~no) based on equations that are expressed at the level of a grid cell. We prefer to call them 'simulators' in order to avoid confusion with other meanings of the word `model'.  

Accounting for physical constraints  is naturally considered to strengthen the reliability of a prediction, especially if this prediction involves a situation that is unprecedented. 
Lorenz's proposal is therefore a generally well accepted way of dealing with the problems of climate predictability at all time scales. However, it has a number of drawbacks that need to be emphasised in the present context.
An often-heard objection to  Lorenz' proposal is computing time. State-of-the-art simulators resolving the dynamics of the ocean and the atmosphere are designed to be run on time scales of the order of a century. One experiment of thousand years with such a simulator may take months of computing time when it is possible. Experiments over several millennia therefore require changes in the simulator design. Modellers have generally chosen to degrade, sometimes very drastically, the resolution of the atmosphere and ocean and simplify or suppress many of the parametrisations in order to gain computing time. Cloud cover, for example, may simply be considered as constant. In the meantime, they attempted to account for the dynamics of other components of the climate system, such as the ice sheets and the carbon cycle, which react on the longer time scales. The simulators satisfying these criteria are known as Earth Models of Intermediate Complexity, or simply EMICS \citep{claussen02emic}. 

EMIC experiments designed to test the early anthropogenic hypothesis were presented soon after the original publication of \citet{ruddiman03}. An EMIC called CLIMBER-SICOPOLIS (\citep{petoukhov00,Calov05aa}) was used to estimate the effects of declining CO$_2$ concentrations during the Holocene from about 265~ppm to about 235~ppm (a similar decline was imposed on methane concentrations). With that scenario, CLIMBER does not accumulate ice any time during the Holocene, while it correctly reproduces the increase  in ice area during the Eemian / Weschelian transition (115~ka ago) \cite{claussen05}. Similar experiments were carried out with the LLN-2D simulator of \cite{gallee92}.  Consistent with the experiments with CLIMBER, reasonable no-antropogenic scenarios with CO$_2$ declining to about 240$~ppm$ cause almost no accumulation before several tens of thousand years in the future.  Only very drastic declines in CO$_2$ concentrations, such as to reach a CO$_2$ concentration of 225~ppm at present and 206~ppm 4~ka in the future, yield some ice accumulation (about 27~m of equivalent eustatic sea-level in 22~ka). 

Results of experiments with EMICS may however be questioned because they misrepresent (if at all)
features of the climate system that may be important to understand and predict its slow evolution, such as monsoons, western boundary currents, modes of variability like ENSO and the details of the ocean circulation in the Southern Ocean. 
For these reasons a number of experiments were published, based on simulators with more state-of-the-art representations of ocean and atmosphere dynamics. These simulators are not designed to compute the accumulation of ice but they are useful to indicate spots where snow is likely to accumulate from one year to the next, and/or to analyse the conditions propitious for accumulation. They also provide information about the sensitivity of the ocean surface and circulation to changes in the astronomical forcing and greenhouse gas concentrations. 

In the present special issue, three articles discuss experiments with the Community Climate Model in different configurations: \citet{Vavrus11aa} examine the influence of topography representation on the snow accumulation process and find that accounting for higher resolution topography increases the sensitivity of snow accumulation on the external forcing; \cite{Kutzbach11aa} examine the influence of decreasing greenhouse gas concentrations on sea-surface temperatures and attempts to quantify the effects on the carbon cycle; they suggest that the feedback of the surface cooling on the carbon cycle is substantial enough to accommodate Ruddiman's suggestion of a natural amplification of the natural perturbation \cite{ruddiman07rge}.  \cite{Vettoretti11aa} uses the ocean-atmosphere version of the Community Climate Model and compares the effects of decreasing CO$_2$ concentrations with those of the orbital forcing on snow accumulation and the abyssal circulation in the Atlantic. They come to a somewhat challenging conclusion for the Early Anthropogenic Hypothesis, that is, astronomical forcing is a more important  driver of ice accumulation than CO$_2$. 
Earlier, \citet{Vettoretti04aa} presented a series of eight experiments with the Canadian Climate Centre Atmospheric Model GCM2 to identify the conditions propitious to year-to-year snow accumulation on ice nucleation sites (see also references to closely related work in that article). They found that  obliquity greatly influences snow accumulation in their model. 

Here, I would like to point to perhaps a more fundamental challenge to Lorenz's vision. 
Even the most sophisticated simulators are an incomplete (truncated) representation of reality. Some of the truncated processes are represented with semi-empirical formulae, called parameterisations (see \cite{Palmer05aa} for a very accessible overview). A fraction of this uncertainty---called the parametric uncertainty---can be quantified by considering the  sensitivity of predictions on parameters involved in the representation of the truncated processes. Pioneering works of \cite{murphy04qump} on this subject lead to great research activity.
However, it will never be possible to guarantee that all the physical and biogeochemical processes are correctly and accurately represented in a numerical model of the climate system. There is structural uncertainty. 
In particular, remember that CO$_2$ is active not only as a greenhouse gas, but also as a fertiliser and an ocean acidifier; changes in CO$_2$ concentrations at glacial-interglacial time scales have innumerable consequences on life, and hence on climate. 

Saltzman already noted  \cite{saltzman02book}(sect. 5.1) that expecting from a climate simulator to reproduce ice ages  without any prior information about the timing of glacial-interglacial cycles is illusory, so huge would the accuracy needed on the snow accumulation imbalance and carbon dioxide fluxes to capture the rate and timing of glacial cycles. 
 Given that glacial cycles cannot be simulated \textit{ab initio}, past observations have to be somehow incorporated into the simulator. 
Very often, the procedure is quite implicit: The uncertain parameters of the climate simulator are varied until the simulated climate (or climate changes) is reasonably realistic. This is called tuning. 
One must be clear about the fact that published predictions of the next glacial inception are all obtained with  climate simulators 'tuned' on past climate observations. For example, the LLN-2D simulator, used in  \citet{Berger2002An-exceptionall}, \cite{loutre03gpc} and  \cite{loutre04epsl} was tuned to capture the timing and rate of the latest glacial inception, about 115~ka ago (Andr\'e Berger, pers. comm.). Confidence in the simulator was gained by the fact that once tuned, it reproduces the last glacial-interglacial cycle \citep{gallee92} reasonably well, as well as the dynamics of previous interglacial periods \citep{loutre03gpc} and, admittedly with more difficulties, the succession of marine isotopic stages 12-11-10 (400,000 years ago) \citep{loutre04epsl}. 

The problem is that tuning a simulator is a fairly non-transparent and non-optimal way of using past climate information for the prediction. 
It is non-transparent in the sense that it is not always clear which information has been incorporated and which one was effectively  predicted (or 'hindcasted'); it is non-optimal in the sense that only a fraction of the palaeoclimate information has been used in the inference process, and that the simulator can never be perfectly calibrated. Furthermore, the `satisfaction' criteria which preside the tuning process are not explicitly defined and the consequences of the deviations between simulator results and observations on predictions are not explicitly quantified. 


\section{Dynamical systems as models of glacial cycles \label{sec:dyn.sys}}
\subsection{What do we learn from simple dynamical systems?}
The complexity and multitude of mechanisms involved in glacial-interglacial cycles may leave the modeller hopeless. Fortunately, this is a fact of nature that very complex systems may exhibit fairly regular and structured dynamics.
This phenomenon, called 'emergence', is related to the fact there are constrains on the system taken as a whole.
The dynamics of the system may therefore be inferred and understood by identifying the constraints that are relevant at the scale of interest without having to consider all the interactions  between the individual components of the system. 

This argument justifies the interest of scientists for `conceptual' models of glacial cycles. These models do have educational interest and they also have predictive power if they are correctly formulated  (see examples in \citet{paillard01rge}; 
for more general background about complex systems consider the monographs of \cite{Haken06aa} and \cite{Nicolis07aa}). 

For illustration consider the following minimal conceptual setting. Suppose that the state of the climate system may be summarised by two variables. Call them $V$ and $D$. Variable $V$ represents the total continental ice volume. It accumulates variations in time due to the astronomical forcing $F(t)$ plus a drift term associated with the variable $D$, which is defined later. Mathematically this reads, if  $\delta V$ is a variation in $V$ over a fixed time interval $\delta t$:
\begin{subequations}
\begin{equation}
\tau \frac{\delta V}{\delta t} = - (D + F(t) + \beta)
\label{v1}
\end{equation}
Parameter $\beta$ is introduced later. $\tau$ is a characteristic time  that controls the rate at which ice volume grows or decays, and which is tuned on observations. 
Variable $D$ represents the state of a climatic component that responds with a hysteresis behaviour to changes in ice volume. The variations  in surface pCO$_2$ associated with ocean circulation changes may present such a behaviour  (e.g. \citep{Gildor01aa, paillard04eps}). This is mathematically modelled as follows:
\begin{equation}
\tau \frac{\delta D}{\delta t} = \alpha (9D^3 - 3D - V + 1),
\label{v2}
\end{equation}
\label{FitzHughNagumo}
\end{subequations}
where $\alpha$ is another parameter that controls the rapidity of the transition between the two branches of the hysteresis
(to understand the principle of this second equation, consider the situation $V=1$. There are then two possible values of $D$ that would be in equilibrium with $V$ (i.e.: $\frac{\delta D}{\delta t}=0$): $D=0.57$ or $D=-0.57$. Whichever is approached depends on the system history, i.e.: there is hysteresis). 
Equations (\ref{FitzHughNagumo}a,b) constitute the dynamical core to several conceptual models of ice ages available in the  literature, including models by \cite{Saltzman91sm,  Gildor01aa} and \cite{paillard04eps}. Mathematicians will recognize a particular formulation of the celebrated  FitzHugh-Nagumo model  (e.g. \citet{Izhikevich:2006}) originally developed to study neuronal responses to electrical stimuli (the term '+1' added at the end of equation \eqref{v2} plays no other role than constraining $V$ to be almost always positive, which is more physical for an ice volume). 
In the absence of astronomical forcing the behaviour of $V$ is mainly dictated by $\beta$. It may converge to a fixed point close to zero (if $\beta\leq1/3$), much above zero ($\beta \geq 1/3$) or  oscillate between the two (if $-1/3<\beta<1/3$). Within the oscillation regime $\beta$ controls the asymmetry of the spontaneous oscillations and $\alpha$, the sharpness of the shifts between the glaciation and deglaciation regimes. Here we chose $\beta=0.2$ to capture the notion of slow glacial build-up and fast deglaciation, and $\alpha=3$ for rapid shifts in $D$ compared to variations in $V$. The oscillations are also known as relaxation oscillations, first documented by Van der Pol in 1921 \citep{Kanamaru:2007}. 

\begin{figure}[ht]
\begin{center}
\includegraphics[width=\textwidth]{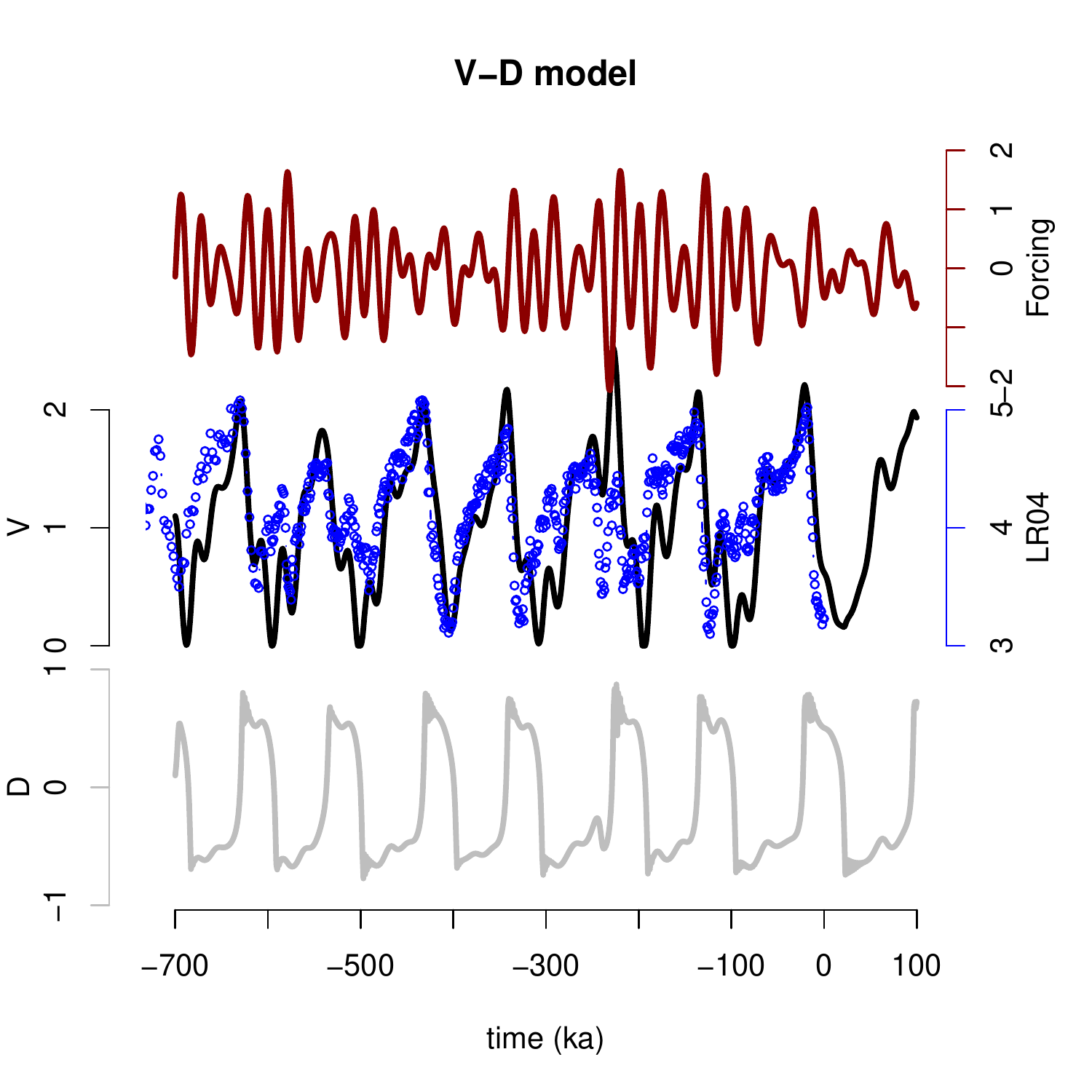}
\end{center}
\caption{Time-series generated with a two-variable model, equations \eqref{v1} and \eqref{v2}. The system is forced by a synthetic mix of obliquity and precession. The variable $V$ integrates the forcing with a correction $D$. $D$ shifts between two states according to a hysteresis behaviour controlled by $V$. In this conceptual model $V$ may be interpreted as the total volume of continental ice, and $D$ is often associated to the state of the ocean circulation. $V$ is here superimposed to the benthic stack by \cite{lisiecki05lr04}, which is  appropriately scaled to highlight the similarity between the simulation and the palaeoclimate record.}
\label{fig:vdmodel}
\end{figure}

For forcing the system we use a linear combination of climatic precession and obliquity  in order to capture the notion that summer insolation controls ice ages ($F(t) = 30[e\sin\varpi + \varepsilon'$],  where $'$ is the departure to the mean). With this scaling the forcing amplitude is approximately of the order of $1$ and thus of the same order as $D$.
There are already two lessons to be drawn from this simple experiment. 
Simulations with this model are shown on Figure \ref{fig:vdmodel}. First, observe that the simulated future climate history  is an ice volume decline up to 20~ka into the future, in spite of the fact that the model correctly reproduces the increases in benthic trends at the end of marine isotopic stages 11 (400~ka ago),  9 (325~ka ago, although the model spuriously produces a subsequent drop in ice volume), 7 (200~ka ago) and 5e (115~ka ago,  but as for stage 9 the subsequent drop towards stage 5c is too strong).  
Hence, the argument that present ice volume should be growing (assuming no anthropogenic perturbation) because  it did so during analogous situations in past interglacials is not sufficient. It is necessary to demonstrate, in addition, that this simple model is inappropriate. 
Second, observe that $D$ shifts to negative values at the end of the deglaciations and, in this model, this shift preconditions glacial inception. This may justify why, rather than considering insolation, it has sometimes been decided to seek an analogue to the present by considering the timing of terminations, because they are a proxy for the state of system variables that cannot be immediately observed. Namely,   \citep{EPICA04} chose to align the previous deglaciation on the deglaciation that lead to stage 11. Based on the observed trends in the $\delta$D, they estimated that glacial inception should not occur before more than ten thousand years from now. This alignment was debated on the ground that it violates the insolation alignment \citep{crucifix06eos,  ruddiman07rge} (further informal debate and correspondence followed). However, they may be a more fundamental reason why any `alignment', i.e., any blind application of the analogues method, may yield a wrong or at least overconfident prediction. 

To see this slightly modify \eqref{v2} to add a random number of standard deviation 0.05 every 1~ka (assume $\sigma$ normally distributed) :
\begin{equation}
\tau \frac{\delta D}{\delta t} = \alpha (D^3/3 - D - V + 1) + 0.05 \sigma \sqrt{\delta t}
\tag{\ref{v2}'} \label{v2p}
\end{equation}
The random (stochastic) process parameterises unpredictable ocean-atmosphere disturbances such as Dansgaard-Oeschger events  (cf. \cite{Ditlevsen09ab}), or volcanic eruptions, which impact on climate dynamics. 
The theory that allows us to represent non-resolved dynamical processes with stochastic processes was introduced in climatology by \citet{hasselmann76}. \citet{Nicolis05aa} discuss to what extent such parameterisations may also be used to estimate the effect of accumulating model errors, 
and a full volume entirely devoted to stochastic parameterisations was recently published \citep{Palmer10aa}. 

\begin{figure}[ht]
\begin{center}
\includegraphics[width=\textwidth]{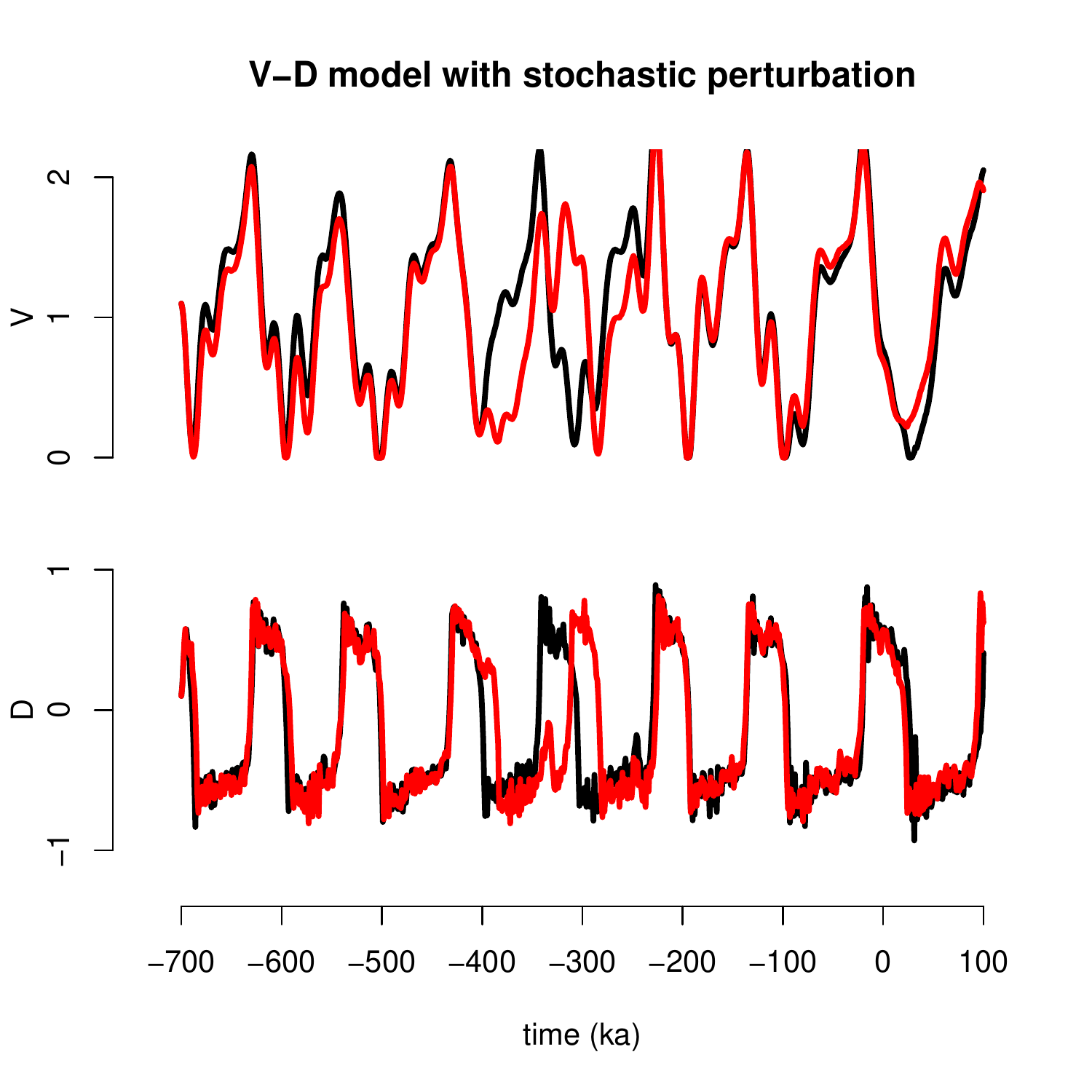}
\end{center}
\caption{
Same model as \figref{fig:vdmodel} but with a small stochastic forcing added to $D$ (equation \ref{v2p}). The two time-series shown correspond to two realisations of the stochastic forcing. Observe the phase shift aroud $-400$\,ka. It corresponds to temporary loss of synchronisation of the system on its (astronomical) forcing.} 
\label{fig:vdstoch}
\end{figure}

Together, equations \eqref{v1} and \eqref{v2p} form a \textit{stochastic dynamical system}.
Two  trajectories  simulated with two different realisations of the random numbers are shown on (\figref{fig:vdstoch}). 
In principle the standard deviation of the noise (0.05) is an uncertain quantity that should  be inferred from realistic computer experiments and / or calibrated on observations. Given that the present purpose is merely illustrative  we are satisfied with the fact that the noise added to the evolution of $V$ and $D$ looks reasonable compared to typical benthic and ice core records. 
Indeed, it is seen that the stochastic disturbances are here too small to compromise the shape of glacial cycles and the action of astronomical forcing as a pace-maker, but they are large enough to alter the timing of glacial events and shift the succession of glacial cycles. 
Observe that the climatic history shown in black is similar to the deterministic simulation shown in  \figref{fig:vdmodel}. The red one undergoes a shift around $-400$\,ka. The length of the simulated interglacials around $-400$\,ka are  differ in the two experiments, in spite of the fact that the preceding deglaciations are in phase and of similar amplitude.

A similar phenomenon was is fact visible in the experiments with the conceptual model of \citet{paillard98}, who showed that two future climate scenarios, one with immediate glacial inception and one with a long-delayed glacial inception, with only tiny changes in model parameters.  It is possible to explain these shifts with arguments of dynamical system theory (article in preparation). They are related to the fact that the non-perturbed dynamical system may accommodate different plausible climate histories, characterised by different series of deglaciation timings. At certain times,  random perturbations (or small parameter variations) may cause a shift from one possible history to another one. The (unsolved) mathematical problem is to identify and characterise the moments at which such phase shifts are the most likely to occur. 
The possibility of such shifts contrast with the earlier conclusions of \citet{tziperman06pacing}. Whether  phase shifts effectively occurred in the true climatic history is still to be elucidated. 
However, if this phenomenon is confirmed, it implies that the predictive horizon of glacial cycles is intrinsically limited and that the analogues method may lead us to be overconfident about what can actually be predicted. 
\subsection{Developing dynamical systems of glacial cycles. }
The above example suggests that the gross dynamics of glacial-interglacial cycles may be characterised by a simple dynamical system.  However, \cite{tziperman06pacing} previously warned us about a form of ambiguity related with the interpretation of such simple dynamical systems of ice ages. Indeed, we have not been specific at all about the meaning of the variable $D$.  
In the model of \citet{paillard04eps}, the role of $D$ is played by a variable quantifying the overturning of the southern ocean circulation. \citet{saltzman90sm} choose to introduce a third variable in order to distinguish the ocean circulation from the carbon cycle. In their model,  the non-linear terms appear in the carbon-dioxide equations. More interpretations of the meaning of $V$ and $D$ are possible.
Ambiguity \textit{per se} is not necessarily a bad thing. It is useful to explore and understand the system dynamics without having to enquire about the details of the mechanisms involved. However, it is desirable to attach a more concrete meaning to climatic variables in order to explore mechanisms and provide credible predictions outside the range of observations. 
There are two complementary ways of doing this. The first one is to verify that the relationships between the different variables are \textit{compatible} with the information inferred from experiments with numerical simulators. For example, eq. \eqref{v1} encodes a linear relationship between the ice mass balance and insolation. How does it compare with climate simulators? The second one is to connect the variables with actual palaeoclimate observations. If $V$ and $D$ represent ice volume and, say, the North Atlantic Overturning cell, their simulated variations should be consistent with the current interpretations of planctonic and benthic foraminifera records. 

\begin{figure}[ht]
\begin{center}
\large\textsf{Strategy to design the 'climate stochastic generator' }
\includegraphics[width=\textwidth]{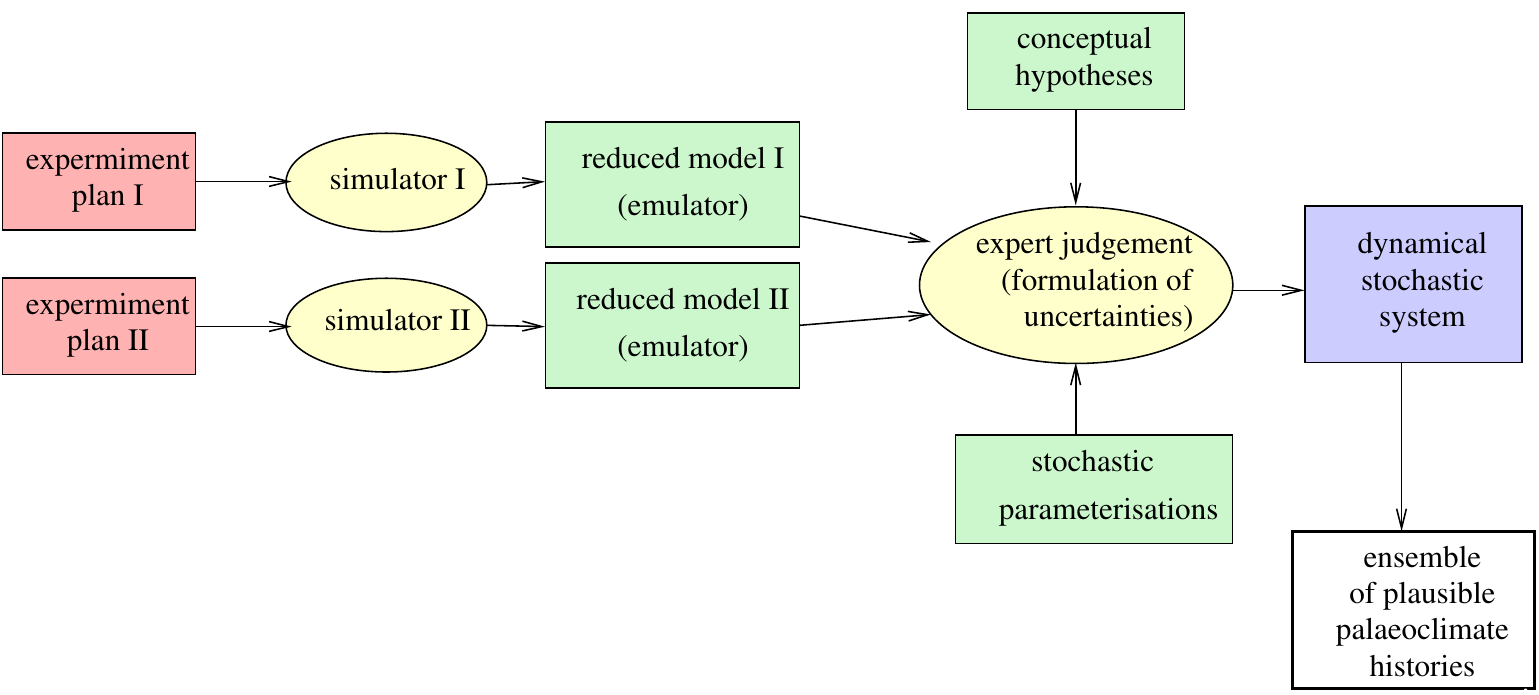}
\end{center}
\caption{
Route map for the design of a stochastic generator of climate histories. Statistical approximations of large numerical simulators (surrogates or emulators) are developed on the basis of appropriate plans of experiments. These mathematical objects are much smaller and tractable than the original simulators, and may be coupled with each other and/or with more conceptual models. Expert elicitation is required for assessing the uncertainties associated with numerical simulators and formulate of prior probability distributions of parameters. The resulting stochastic dynamical system may then be used to generate ensembles of palaeoclimate histories compatible with this model.
}
\label{fig:csim}
\end{figure}

In this section we focus on how we include information inferred from numerical simulators in simple dynamical systems (\figref{fig:csim}  supports the expos\'e). Our ambition is to reach a level of understanding of the simulator response that is more general than mere predictions of the future ice volume evolution obtained with this simulator. 
To this end we propose to examine the \emph{bifurcation structure} of the simulator in response to varying input parameters.

To this end, consider again LLN-2D. This model is an implementation of equations of quasi-geostrophic motion in the atmosphere and rheological equations for ice-sheet dynamics assembled with the purpose of studying glacial cycles.  
As we noted above the level of sophistication of LLN-2D is largely superseded by modern simulators. Nevertheless it is a valid example because LLN-2D was used in the debate over the early anthropogenic  \citep{crucifix05anthropocene,loutre07interglacials} .

\begin{figure}[ht]
\begin{center}
\includegraphics{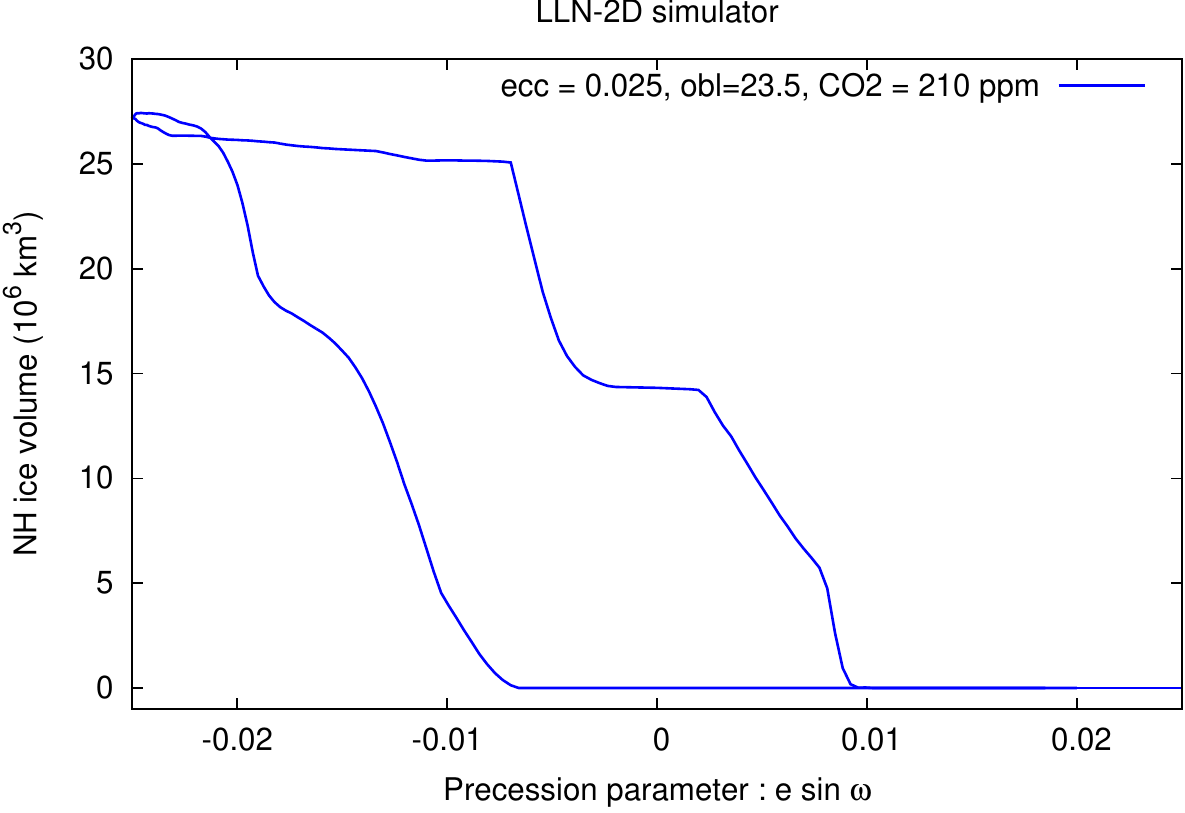}
\end{center}
\caption{Simulated volume of ice over the northern hemisphere near steady-state as a function of precession using the LLN-2D simulator by \cite{gallee92}, and given fixed eccentricity, obliquity and CO$_2$ concentration. The experiment consists in varying the precession parameter slowly all along the trigonometric cycle in 400~ka of model time. 
\label{fig:llnhys}}
\end{figure}

In this example, we attempt to determine the number and stability of steady-states obtained with LLN-2D, as a function of the precession parameter. Mathematicians  devised very ingenious and sophisticated ways of exploring the dynamical structure of large mathematical systems but here we follow a very simple and intuitive method, previously used to analyse simulators of the ocean-atmosphere  \cite{Rahmstorf05aa} and ocean-atmosphere-ice-sheet systems  \cite{Calov05aa}. It  is known as the `hysteresis experiment'. This is achieved as follows: consider eccentricity of 0.025, obliquity of $23.5^\circ$, a CO$_2$ concentration of 210~ppm, and run the simulator until equilibrium with $\varpi=90^\circ$ (perihelion on the 21st June). Then, slowly vary $\varpi$ along the trigonometrical circle (the experiment takes the model equivalent of 400,000 years) and record the total continental ice volume as $\varpi$ varies. The resulting curve (\figref{fig:llnhys}) depicts an ensemble of quasi-equilibrium points. The first observation is that there is at least one stable equilibrium response for all the forcings in the range explored. Secondly, there are two stable states for $e\sin(\varpi)$ within about $-0.02$ and $+0.01$. 
A similar curve was calculated for a CO$_2$ concentration of 260~ppm. It is similar in shape but the two-stable-state regime now ranges from $e\sin(\varpi)=-0.24$ to $0$ and the maximum ice volume is $16\ 10^6~\mathrm{km}^3$). 
Similar results were previously obtained with the CLIMBER-SICOPOLIS simulator \citep{Calov05aa}. 

The bifurcation structure that emerges from  LLN-2D was not apparent until we did the simulations. We have therefore learned a new bit of information that we are ready to trust, or at least to test against the observations. 
This information may be encoded as a simple equation, which is known to have a similar bifurcation structure as the one found in LLN-2D:
\begin{equation}
\mathrm{\delta }V = -\phi_V(V,\mathrm{CO}_2, \varepsilon, e,\varpi,\Psi) \delta t
\label{eq:llnemu}
\end{equation}
where $\delta V$ is a variation in ice volume over a time-step $\delta t$, $\phi_V$ is a function of ice volume, astronomical forcing and CO$_2$ and other parameters (summarised by $\Psi$) (the full expression is given in Appendix 1). 

\begin{figure}[ht] 
\begin{center}
\includegraphics[width=\textwidth]{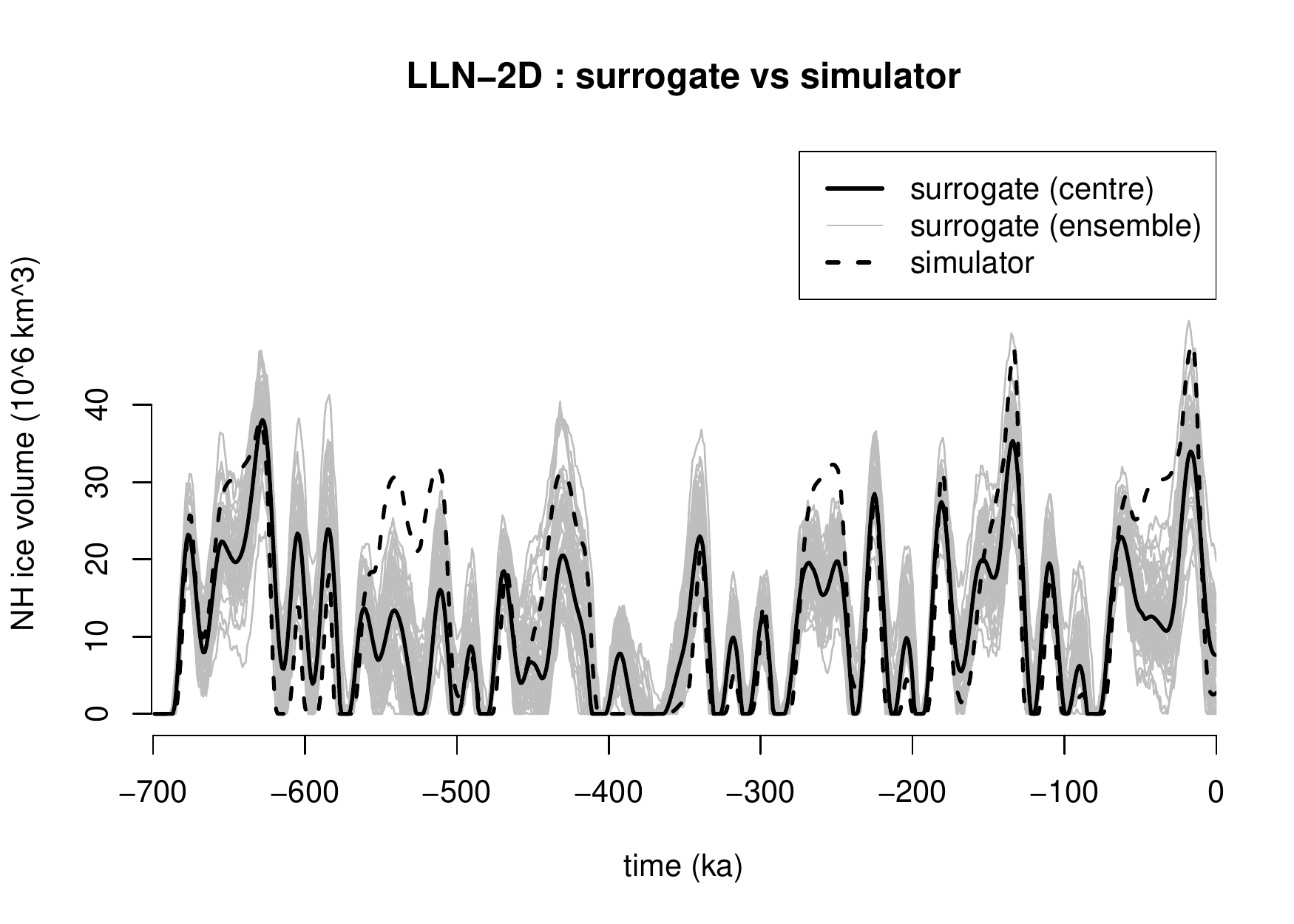}
\end{center}
\caption{Simulation of ice volume with the LLN-2D model (red) forced by astronomical forcing and variations in CO$_2$ \citep{petit99,Luethi08aa}, compared with 50 realisations of the emulator (eq. \eqref{eq:llnemu}). The blue trajectory is obtained with the emulator using the central estimates of regression coefficients. All experiments are deterministic. }
\label{fig:lln-emulator}
\end{figure}

Equation \ref{eq:llnemu} is a \textit{surrogate} of the LLN-2D model: it is a simple, fast process that behaves dynamically  in a way that is similar to LLN-2D. Here, we calibrated the parameter $\Psi$ such that the surrogate reproduces approximately the behaviour of LLN-2D over the last 800~ka. The calibration did not yield a simple \textit{best} value of $\Psi$, but rather a distribution of plausible values of the parameter $\Psi$ that reproduce reasonably well the model output. The resulting fit is shown on \figref{fig:lln-emulator}.

By finding a surrogate for LLN-2D  we have progressed on two fronts. 
First, we have distilled the vast network of partial differential equations implemented in LLN-2D into a much simpler structure. This structure can now be compared with canonical dynamical systems in order to comprehend its important dynamical characteristics (\cite{Ditlevsen09aa} precisely studied an equation similar to \eqref{eq:llnemu} in the context of glacial cycles). Secondly, it is possible to efficiently express our beliefs and uncertainties about this simulator. Namely, uncertainty about the position of tipping points may be expressed as uncertainty on the coefficients $\beta_i$  (cf. Appendix 1). 

Unfortunately the LLN-2D is a very rough simulator by today's standards. One should therefore attempt to correct or augment eq. \eqref{eq:llnemu} based on the knowledge gained from simulators more specifically designed to study the response of the atmosphere and oceans to changes in forcing and boundary conditions. 
This is where well-designed experiments with general circulation simulators of the ocean and the atmosphere, such as those reviewed in the previous section, may be useful. 
Such plans of experiments may help us to locate points in the parameter space from which snow may accumulate from one year to the next. 
More generally, they should allow us to explore the (possibly non-linear) response structure of the atmosphere-ocean system to changes in forcing and boundary conditions.
To this end, statisticians have developed two concepts that should be useful:
\begin{description}
\item[The experiment design theory.] This theory provides practical guidance to 
 efficiently sample the space of possible inputs (forcing and parameters) of a numerical model in order to learn as much as possible about the simulator with as few experiments as possible  \citep{santner03}.
\item[The Emulator ]  is an interpolator  that predicts the response of the climate simulator for other parameters than the experiments actually made. It also provides an uncertainty of the error associated with the interpolation process \citep{ Kennedy01bayes,Rougier09aa}
\end{description}

Just as the LLN-2D surrogate, the emulator provides us with a fast, efficient, high-level description of the general circulation simulator response. It may constitute the basis for non-linear parameterisations of, the  northern hemisphere snow accumulation balance response to obliquity, precession, CO$_2$ and ice volume. In turn, these parameterisations can be included into a  dynamical model of ice ages. 


 At last,  it happens that certain processes supposed to relevant at the glacial time scales are difficult to simulate reliably with numerical simulators. For example, calculating the response of the ocean circulation near the Antarctic ice shelves to changes in sea-level is particularly challenging because it involves connexions between processes at very different spacial scales. Yet, \cite{paillard04eps} speculated that these circulation changes may play a role on carbon cycle dynamics in glacial cycles, and they support their claim with a number qualitative arguments. Likewise, \cite{Ruddiman06ab} suggests a direct effect of precession on the carbon cycle through tropical dynamics and, possible, southern ocean temperatures. In such cases where numerical simulation are difficult or considered to be too unreliable, the framework of the climate stochastic dynamical model  gives us the freedom to formulate hypotheses about the role of such processes in the form of simple equations, and then to test them against observations.


\subsection{A statistical framework to accommodate observations}
 Many of the ideas expressed above were previously formulated by Barry  \cite{saltzman02book}.
The present programme for investigation and prediction of ice ages proposed here goes further. It follows a  \textit{Bayesian} methodology because  uncertainties are expressed by means  of probability distribution functions and they are updated using observations. 

\citet{Haslett06aa} depicts the problem quite efficiently: the system defined by eq. \eqref{v1}, \eqref{v2p}  may be interpreted as a generator of \textit{a priori} plausible climate histories. They are \textit{a priori} plausible in the sense that they are generated from a system of equations and probability distributions of parameters that express, transparently, our beliefs about the dynamics of the climate system. 

In practice,  the calibration of this `generator' consists in rating these climate histories against palaeoclimate observations. This operation consists in  attributing more or less weight to the different possible model parameters, depending on the likelihood of the climate histories that they generate. This weighting process constitutes the mechanism by which the \textit{a priori} uncertainty on model parameters is \textit{updated} by accounting for observations. 
The more general principle of \textit{model selection} consists, in the presence of  two different models, in deciding which one produces the climatic histories that are the most likely given the observations. 
The concept of Bayesian model calibration was first applied to the problem of glacial cycle dynamics by \citet{hargreaves02}. 

\begin{figure}[ht]
\begin{center}
\large\textsf{Bayesian inference network}
\includegraphics[width=\textwidth]{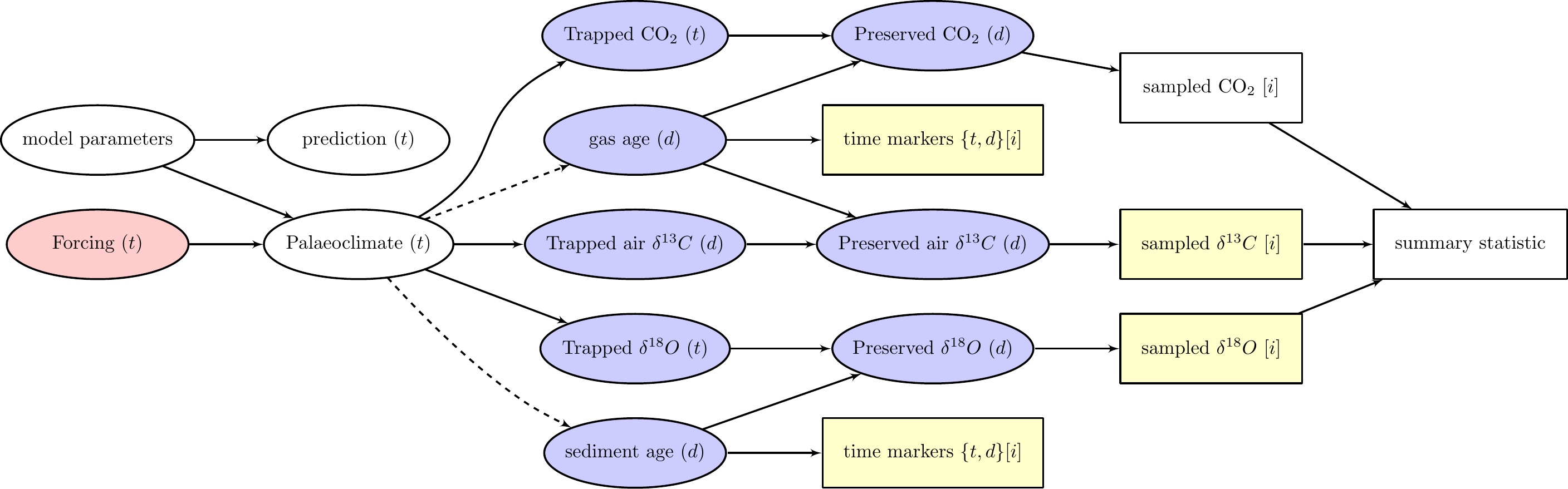}
\end{center}
\caption{Simplified Bayesian inference network for selection and calibration of stochastic dynamical systems of palaeoclimates (cf. \citet{Haslett06aa} for more theoretical background). 
Rectangles correspond to observations or certain quantities. Ellipses are  uncertain quantities. 
Continuous functions of time are symbolically marked with $(t)$;  functions of depth by $(d)$ and discrete series are marked with  $[i]$.
The sequence from left to right corresponds to forcing, climate model, natural archiving of the climate information (blue), sampling and measure (yellow). 
Modelling consists in defining the  relationships symbolised by the arrows and attributing prior probability distributions to uncertain quantities at the top of the chain (e.g.: model parameters). Model validation and selection are judged based on whether actual observations are compatible the model (measured by the likelihood);  statistical calibration consists in propagating the information backwards, from the observations to parameters in order to make predictions (green).  'Summary statistics' are variables such as spectral power or other integrated measures which summarise several observations. }
\label{fig:modelproxy}
\end{figure}

The \textit{likelihood} is, in that framework, an important quantity. It is formally defined as \textit{the probability of observing what was observed, if the simulated palaeoclimate history was correct}. Estimating the likelihood therefore requires to express carefully the connections between  modelled climate histories (the 'model'), the real world,  and observations.  These connections may be visualised by means of a Bayesian graphical network, an illustrative  example of which is given in \figref{fig:modelproxy}. The graphic is greatly inspired from that published by \citet{Haslett06aa} but it is here adapted to the glacial cycles problem. 
Arrows indicate causal dependencies.  Namely, our knowledge of benthic $\delta^{18}O$ sampled at a certain depth depends on the  $\delta^{18}O$ that was actually preserved at that depth (after, e.g., bioturbation), which itself depends on the age corresponding to that depth, as well as on the   $\delta^{18}O$  trapped by the foraminifera at the time corresponding to that age. Further up, 
the generated climate histories  depend on the parameters of the climate stochastic model and on the external forcing. 

The climate stochastic dynamical system used to generate `palaeoclimate histories' is  thus  part of a large framework generating   potential 'palaeoclimate records', chronologies and sedimentation histories.
 \cite{Haslett08aa} provide an algorithm to generate sediment chronologies compatible with time markers (radiocarbon dates or, in our case, tephra layers and/or magnetic inversions). 

 \figref{fig:modelproxy} is  a simplified diagram; the full  network should also include parameters controlling the response of proxies on climate. Nonetheless, it is good enough to clarify some of our ideas about palaeoclimate how inferences are drawn. For example, palaeoclimate scientists often take advantage of astronomical forcing to constrain the chronology of climate records \cite{imbrie84}. The Bayesian graphical network shown here makes it clear that this operation \textit{requires} a climate model (the white ellipse representing climate histories generated by the climate model lies between the forcing and observations).

Bayesian methodologies have become increasingly popular in climate science because they provide transparent ways of expressing our uncertainties, modelling choices \citep{annan06multiple,Rougier2007Probabilistic-i, Rougier07aa, Sanso08aa}
and the distance estimated between the model and reality (on this specific point see \citet{Goldstein09aa}). 

Calibrating dynamical systems is however a very difficult problem. 
In general, the number of possible climate histories compatible with a model is so big that fairly complex algorithms are needed to reduce as much as possible the number of experiments required to efficiently update prior distributions. 
Palaeoclimates  present additional specific challenges  related to dating uncertainties and the complexity of the environmental factors affecting palaeoclimate archives. 
\citet{hargreaves02} proposed a straightforward application of an algorithm well known by statisticians called `Markov-Chain Metropolis Hastings'. The algorithm works well as long as the calibration period is relatively short and that climate trajectories all cluster around a same common history. Unfortunately this  may not be the case if local instabilities occur, as in the example shown on \figref{fig:vdstoch}. Furthermore, the algorithm demands large computing resources. 
\cite{Crucifix09aa} then proposed a solution based on a statistical algorithm called  `particle filter for parameter and state estimation' \citep{lw01}. This is a sequential filter: it `sweeps' observations forward in time from the past until the present in order to generate a posteriori distributions of parameters and model states at the end of the run. The filter seemed to be a powerful solution to the problem posed by the existence of local instabilities. 
Unfortunately (again), further sensitivity studies performed after publication lead us to tone down our enthusiasm. The filter performs poorly on the model selection problem because it fails to discriminate models on the basis of their long-term dynamics. For example, some of the models selected by the filter no longer exhibit glacial-interglacial cycles \citetext{unpublished results; available on request to the author}. 

Hope resides in more advanced Bayesian methods,
which combine the Metropolis-Hastings strategy with the particle filter  \citep{Andrieu10aa}. 
An alternative solution  is to calibrate the parameters on the basis of invariant summary statistics \cite{Wood10aa} using a method known an 'Approximate Bayesian Computation' \cite{Sisson07aa, Wilkinson08aa}. Such statistics allow one  to characterise a climatic trajectory in a way that is not sensitive to its initial conditions, nor to  the exact timing of climatic events. 
For example, one may attempt to calibrate the dynamical system based on the average duration of a glacial cycle, or in the 
`skewness' or `asymmetry' of these cycles \citep{King1996247}.  More complex metrics may be envisaged to capture complex features associated with the non-linear dynamics of the system \citep[and. ref. therein]{Marwan20094246, citeulike:753023}. 
The use of summary statistics, however, should still respect the inference process outlined on \figref{fig:modelproxy}. An easy mistake is  to apply summary statistics on climate model outputs and compare them straightforwardly with palaeoclimate data, without regard to the data preservation, sampling and possibly interpolation processes. Good summary statistics should allow us to efficiently discriminate between climate models and, at the same time, they should be robust to hypotheses about the data preservation and sampling processes. The works by \cite{Mudelsee94aa, Mudelsee01aa,npg-16-43-2009} and ref. therein constitute important references in this respect. 

\section{Conclusion}
The Pleistocene record suggests that modern insolation is not much above that needed for glacial inception. 
However, the complexity of the climate response to the complex astronomical forcing implies that further theoretical elaborations are needed  to transform this statement into a reliable prediction about the slow evolution of climate like: `greenhouse gas should have declined during the Holocene, leading to glacial inception'. 
In particular, there is no perfect insolation analogue to the Holocene in the past and, would there be one, it cannot be guaranteed that climate would behave exactly the same way as during that hypothetical analogue. We mentioned several reasons for this. One is the possibility that  small disturbances may, at strategic times,  delay glacial events by several thousands of years. Locating such 'high-sensitivity' times is an attractive challenge. If the Holocene was such a time,  any small disturbance, including the pre-industrial anthropogenic activity, may have durably inflected the course of climate. 

The dynamical system approach proposed here offers opportunities to tackle this problem in the tradition of hypothetic-deductive scientific investigation, subject to the usual criteria of model parsimony, prediction skill and accommodation of available observations.
We have seen that very simple dynamical systems may already roughly reproduce the benthic curve; on the other hand, the useful output of more sophisticated numerical simulators may also be captured by fairly simply structured surrogates or emulators. These considerations
 encourage our quest for fairly compact and efficient predictive models of glacial-interglacial cycles.   They are our main hope to credibly accommodate the fairly low amount of information present in palaeoclimate records with the flow of information generated by the complex simulators of the climate system. 
Indeed, reduced dynamical models designed to emulate complex numerical simulators contain a much smaller number of adjustable parameters and they are way more computationally efficient.  They are therefore  easier to calibrate  on palaeoclimate observations. Furthermore, they make it possible to infer very general statements about climate predictability based on the large body of literature available on dynamical system theory. 

Nevertheless, one should not be deceived by the apparent success our simple model and its close parents.  Finding a dynamical system that convincingly accommodates  more than one palaeoclimate data series is not trivial.
 A convincing model of glacial cycles should arguably accommodate observations about the oxygen isotopic composition of foraminifera, greenhouse gases, and carbon isotopic data, both in the oceans and in the atmosphere. It should explain spectral peaks, stochastic aspects, the asymmetry of cycles and slow changes in the system dynamics like those  that seemingly occurred over the last million years.  So far, this target  
``is still elusive'' \citep{Raymo08aa}.  


Bayesian inference constitutes a powerful framework to handle knowledge and uncertainties associated with palaeoclimate dynamics. This framework will lead us to formulate probabilistic statements about the next glacial inception  and, more generally, assist our investigations about the climate system mechanics and dynamics. 

Such a programme for statistical-dynamical modelling of (palaeo-)climates is in its beginnings and serious challenges are faced. The more general problem of statistical calibration of dynamical systems is already quite difficult. Palaeoclimate data bear further difficulties, both at the climate system level (intertwining of dynamical structures at different time scales like Dansgaard-Oeschger events and glacial cycles) and the observation level (time uncertainties, sediment disturbances, uncertain relationships between climate state and proxy\ldots)
It is therefore my opinion that we can't yet express our predictions about natural trends in ice volume and greenhouse gases by probabilistic statements, and hence, that natural archives of previous interglacials do not falsify natural explanations for increasing greenhouse gas trends during the Holocene. 

\section*{Appendix : the LLN-2D surrogate}
The form adopted for the surrogate is:
\begin{equation}
\delta V = 
  -[ \beta_0 + \beta_1 (V)(V-\beta_2) +  \beta_\Pi(e\sin\varpi) + \beta_\varepsilon(\varepsilon) +   \beta_C\ln C]\delta t 
\label{LLN_emulator} 
\end{equation}
where $V$ is ice volume, $\delta V$ the ice volume increment over $\delta t = 1$~ka.; $C$ is the CO$_2$ concentration, $\epsilon$ is a random number with Gaussian distribution, $\beta_i$ are regression coefficients. $V$ is further forced to be positive by imposing a lower bound of $-V\delta t$ on  $\delta V$.

Priors on regression coefficients are normal as follows (volumes are in $10^6~\mathrm{km}^3$, angles in radians, carbon dioxide concentrations in ppm and times in ka):~$\beta_0\sim\mathcal{N}(19,3^2)$, $\beta_1\sim\mathcal{N}(1.4,0.4^2)/10^3$, $\beta_1\sim\mathcal{N}(27,3^2)$, $\beta_\Pi\sim\mathcal{N}(60,5^2)$, $\beta_\varepsilon\sim\mathcal{N}(72,5^2)$, $\beta_C\sim\mathcal{N}(3,1^2)$ (conventionally, $\mathcal{N}(\mu,\sigma^2)$ is a normal distribution of centre $\mu$ and standard deviation $\sigma$). Posteriors are estimated by running a particle filter for state and parameter estimation \citep{lw01} on an experiment spanning the interval $[-800~\mathrm{ka};0]$ forced by astronomical forcing \citep{berger78} and CO$_2$ data by
EPICA \citep{Luethi08aa} and Vostok \citep{petit99}. The likelihood function used to run the filter assumes that the errors of the surrogate every 1-kyr time step are independent and normally distributed with a standard deviation equal to $1$. 

The red curve on  \figref{fig:lln-emulator} is a climate simulation with the LLN-2D model, while the blue curve is computed  using the surrogate with the central estimates for $\beta_i$. Grey curves are trajectories obtained with parameters sampled from the posterior distribution. 
Note that the timing of the deglaciations is here constrained by the history of CO$_2$ imposed in these experiments. 
\section*{Acknowledgements}
Material present in this paper was greatly influenced by discussions at the SUPRAnet workshops organised  by Caitlin Buck (University of Sheffield), in particular with Jonty Rougier, John Haslett and Andrew Parnell, as well as by further discussions with Jonty Rougier made possible with the exchange programme between the French Community of Belgium and the British council. The editor (Bill Ruddiman) and two anonymous reviewers provided particularly constructive comments. The author is funded by the Belgian Fund for Scientific Research. This is a contribution to the European Research Council starting grant `Integrating Theory and Observations over the Pleistocene' nr ERC-2009-Stg 239604-ITOP.
\bibliography{/Users/crucifix/Documents/BibDesk.bib}
\listoffigures

\end{document}